\documentclass[aps,pre,twocolumn]{revtex4-1}
\usepackage{graphicx}
\usepackage{amsmath}
\usepackage{mathrsfs}
\usepackage{comment}
\usepackage{color}
\begin{document}

\title{Fractality and degree correlations in scale-free networks}
\author{Yuka Fujiki}
\email{y-fujiki@eng.hokudai.ac.jp} \affiliation{Department of
Applied Physics, Hokkaido University, Sapporo 060-8628, Japan}
\author{Shogo Mizutaka}
\email{mizutaka@ism.ac.jp} \affiliation{School of Statistical
Thinking, The Institute of Statistical Mathematics, Tachikawa
190-8562, Japan}
\author{Kousuke Yakubo}
\email{yakubo@eng.hokudai.ac.jp} \affiliation{Division of
Applied Physics, Hokkaido University, Sapporo 060-8628, Japan}

\date{\today}
\begin{abstract}
Fractal scale-free networks are empirically known to exhibit
disassortative degree mixing. It is, however, not obvious
whether a negative degree correlation between nearest neighbor
nodes makes a scale-free network fractal. Here we examine the
possibility that disassortativity in complex networks is the
origin of fractality. To this end, maximally disassortative
(MD) networks are prepared by rewiring edges while keeping the
degree sequence of an initial uncorrelated scale-free network
that is guaranteed to become fractal by rewiring edges. Our
results show that most of MD networks with different topologies
are not fractal, which demonstrates that disassortativity does
not cause the fractal property of networks. In addition, we
suggest that fractality of scale-free networks requires a
long-range repulsive correlation in similar degrees.
\end{abstract}

\pacs{89.75.Fb, 89.75.Hc, 05.45.Df}

\maketitle

\section{INTRODUCTION}
\label{sec:1}
Networks describing complex systems in the real world are quite
inhomogeneous and complicated
\cite{Albert02,Dorogovtsev08,PasterSatorras15}. The number of
edges from a node, namely degree, for example, is widely
distributed in a network. In fact, many real-world complex
networks have asymptotically power-law degree distributions,
which is called the scale-free property \cite{Barabasi99}. In
addition, the degrees of adjacent nodes via an edge are usually
correlated. Such nearest neighbor degree correlations can be
described by the joint probability $P(k,k')$  of a randomly
chosen edge connecting two nodes with degrees $k$ and $k'$. It
has been empirically known that nodes in social networks tend
to be connected to nodes with similar degrees (assortative
mixing) while technological or biological networks show the
opposite tendency (disassortative mixing)
\cite{Newman02,Newman03}. Network complexity is also
characterized by the shortest path distance (number of edges
along the shortest path) between nodes. From this point of
view, most of real-world networks can be classified into two
classes \cite{Kawasaki10}, namely, small-world networks
\cite{Watts98} and fractal networks \cite{Song05}. In a
small-world network, the shortest path distance averaged over
all node pairs increases logarithmically (or more slowly) with
the total number of nodes. In other words, the minimum number
of subgraphs $N_{\text{B}}$ covering the entire network
decreases exponentially (or faster) with the subgraph diameter,
that is,
\begin{equation}
N_{\text{B}}(l_{\text{B}})\propto e^{-l_{\text{B}}/l_{0}},
\label{eq:def_sw}
\end{equation}
where $l_{\text{B}}$ is the maximum distance between any nodes
in the subgraph (subgraph diameter) and $l_{0}$ is a constant.
It has been shown that many real-world networks possess the
small-world property \cite{Albert02,Watts98}. For a fractal
network, on the other hand, the number of covering subgraphs
decreases with $l_{\text{B}}$ in a power-law manner, i.e.,
\begin{equation}
N_{\text{B}}(l_{\text{B}})\propto l_{\text{B}}^{-D},
\label{eq:def_fr}
\end{equation}
where $D$ is the fractal dimension of the network. This
relation has been observed also in a diverse range of networks
from the World Wide Web to some kinds of cellular networks
\cite{Kawasaki10,Song05,Gallos08,Gallos12}.

The origin of the fractal property in complex networks is still
an open question \cite{Watanabe15}, though the mechanism of the
small-world network formation has been found in the existence
of short-cut edges. It has, however, been empirically
demonstrated that real-world and synthetic fractal scale-free
networks have \textit{disassortative} degree correlations
\cite{Yook05,Song06,Rozenfeld07,Kim09}. If the converse is also
true, that is, disassortative mixing makes a scale-free network
fractal, the origin of fractality would be found in the degree
correlation. Disassortativity is a local property
characterizing nearest neighbor degree correlations, while
fractality is a consequence of long-range structural
correlations. This fact seems to deny the above possibility. It
is, however, known that if edges in a scale-free network are
rewired to maximize or minimize the assortativity $r$ while
keeping the degree sequence (and hence the degree distribution)
then the maximally assortative or disassortative network
becomes to possess a long-range correlation
\cite{Menche10}. The assortativity $r$ is the Pearson's
correlation coefficient for degrees and defined by
\cite{Newman02,Newman03}
\begin{equation}
r=\frac{\displaystyle 2\sum_{(i,j)\in E}k_{i}k_{j}-\frac{1}{2M}\left[\sum_{(i,j)\in E}(k_{i}+k_{j})\right]^{2}}
  {\displaystyle \sum_{(i,j)\in E}(k_{i}^{2}+k_{j}^{2})-\frac{1}{2M}\left[\sum_{(i,j)\in E}(k_{i}+k_{j})\right]^{2}} ,
\label{eq_assortativity}
\end{equation}
where $k_{i}$ is the degree of node $i$, $M$ is the total
number of edges, and $E$ is the set of undirected edges in the
network. A maximally assortative network has an onionlike
structure consisting of communities of regular subgraphs
\cite{Menche10,Schneider11}. Such a network displays a
long-range structural correlation in which maximum degree nodes
are distant from minimum degree ones. A maximally
disassortative (MD) network also shows a community structure.
In each community, all nodes with a specific low degree and
higher degree nodes are connected alternately \cite{Menche10}.
Thus, a minimum (or maximum) degree node is located far away
from an intermediate degree node in the MD network, which
implies a long-range correlation. Therefore, there is a
possibility that even short-range disassortative degree mixing
induces the long-range fractal correlation, as pointed out by
Ref.~\cite{Kim09} through the analysis of random critical
branching trees. Furthermore, considering the entropic origin
of disassortativity \cite{Johnson10}, we can discuss the
formation mechanism of fractal scale-free networks in
connection with a process maximizing the entropy.

In this paper, we examine the possibility that a negative
degree correlation between nearest neighbor nodes makes a
scale-free network fractal. To this end, we first prepare an
uncorrelated scale-free network that is guaranteed to become
fractal by rewiring edges. Then, this network is again rewired
so that the nearest neighbor degree correlation becomes
maximally negative. If the rewired MD network exhibits the
fractal nature, one can conclude that fractality of scale-free
networks is induced by disassortative degree mixing. Our
results, however, show that disassortativity does not always
make scale-free networks fractal. We also suggest that
fractality seems to require a long-range repulsive correlation
between similar degree nodes.

The rest of this paper is organized as follows. In
Sec.~\ref{sec:2}, we explain how to prepare the initial network
$G_{\text{ini}}$ that can be fractal scale free by rewiring
edges. The rewiring method to obtain MD networks is also
described in this section. Our results on the fractal property
of rewired disassortative networks are presented in
Sec.~\ref{sec:3}. A long-range degree correlation is also
argued. Section \ref{sec:4} is devoted to the summary and
remarks.

\section{PREPARATION OF NETWORKS}
\label{sec:2}
We investigate the fractal property of MD networks formed by
rewiring edges while keeping the degree distribution of a given
initial uncorrelated scale-free network $G_{\text{ini}}$. In
this section, we explain how to prepare the initial network
$G_{\text{ini}}$ and how to rewire it so as to realize a MD
network.

\subsection{Preparing the initial network}
\label{sec:2_1}
\begin{figure}[tttt]
\begin{center}
\includegraphics[width=0.48\textwidth]{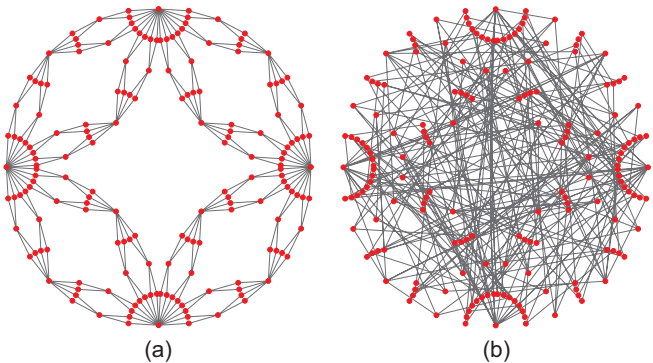}
\caption{(a) The $4$th generation $(2,2)$-flower
with $N=172$ nodes and $M=256$ edges. (b) A network formed by
$5M$ operations of RR for the network shown by (a).}
\label{fig:1}
\end{center}
\end{figure}

The initial uncorrelated scale-free network $G_{\text{ini}}$
must be guaranteed to become fractal by rewiring edges. Such a
network can be constructed by rewiring randomly edges of an
original scale-free fractal network $G_{0}$. We adopt the
$(u,v)$-flower with $2\le u\le v$ as the network $G_{0}$
\cite{Rozenfeld07,Dorogovtsev02}. In the $(u,v)$-flower model, we
start with the cycle graph consisting of $w\equiv u+v$ nodes
and edges [the first generation $(u,v)$-flower]. The $n$th
generation $(u,v)$-flower $G_{n}^{(u,v)}$ is obtained by
replacing each edge in $G_{n-1}^{(u,v)}$ by two parallel paths
of $u$ and $v$ edges. The number of nodes $N$ and the number of
edges $M$ in $G_{n}^{(u,v)}$ are given by
\begin{eqnarray}
N&=&\frac{(w-2)w^{n}+w}{w-1} , \label{uv_N} \\
M&=& w^{n} .
\label{uv_M}
\end{eqnarray}
The degree of a node in this network falls into any of $2^{m}$
with $m=1,2,\dots,n$, and the number of nodes $N_{k_{m}}$ with
the degree $k_{m}=2^{m}$ is given by
\begin{equation}
N_{k_{m}}=\begin{cases}
(w-2)w^{n-m} & (1\le m<n) ,\\
w & (m=n),
\end{cases}
\label{uv_Nkm}
\end{equation}
which specifies the degree sequence of the network. Thus, the
degree distribution function $P(k)$ for $k\gg 1$ of
$G_{n}^{(u,v)}$ with large $n$ is proportional to
$k^{-\gamma}$, where
\begin{equation}
\gamma=1+\frac{\log w}{\log 2}.
\label{uv_gamma}
\end{equation}
The network with $u\ge 2$ exhibits fractality with the fractal
dimension \cite{Rozenfeld07},
\begin{equation}
D_{\text{f}}=\frac{\log w}{\log u}.
\label{uv_Df}
\end{equation}
The $4$th generation $(2,2)$-flower with $172$ nodes and $256$
edges is depicted in Fig.~\ref{fig:1}(a).

The initial network $G_{\text{ini}}$ is formed by rewiring
randomly edges in the scale-free fractal $(u,v)$-flower as
$G_{0}$. The random rewiring (RR) procedure is performed as
follows:

(i) Choose randomly two edges $(i_{1},j_{1})$ and
$(i_{2},j_{2})$ with four different end nodes, namely
$(i_{1}-i_{2})(i_{1}-j_{2})(j_{1}-i_{2})(j_{1}-j_{2})\ne 0$.

(ii) Rewire these edges to $(i_{1},j_{2})$ and $(i_{2},j_{1})$,
if this re\-wiring process does not make multiple edges.

(iii) Repeat (i) and (ii) enough times.

Since the above rewiring preserves the degree of each node, the
degree sequence (and thus the degree distribution) of the
network after RR's does not change from the original one. If
the number of repetition times is much larger than the number
of edges $M$, the rewired network $G_{\text{ini}}$ possesses
essentially the same statistical properties as the
configuration model \cite{Bekessy72,Molloy95,Newman01} with the
degree sequence specified by Eq.~(\ref{uv_Nkm}). Therefore,
$G_{\text{ini}}$ has no degree correlations, as well as no
structural correlations such as fractality. It should be,
however, emphasized that the network $G_{\text{ini}}$ is
guaranteed to become fractal by rewiring edges. As an example
of $G_{\text{ini}}$, the network shown in Fig.~\ref{fig:1}(b)
is formed by the RR procedure (i)-(iii) from the $(2,2)$-flower
of Fig.~\ref{fig:1}(a).

\subsection{Maximally disassortative network}
\label{sec:2_2}
In order to make MD networks $G_{\text{MD}}$ from an initial
network $G_{\text{ini}}$, we first quantify the degree of
assortative mixing in a network. The assortativity $r$ defined
by Eq.~(\ref{eq_assortativity}) is widely used for this
purpose. If $r$ is positive, the network is assortatively mixed
on degrees of nodes, while a network with a negative $r$ shows
disassortative mixing. It has been, however, pointed out that
$r$ strongly depends on the network size and cannot be negative
for infinitely large scale-free networks if the degree
distribution decays asymptotically more slowly than $k^{-4}$
\cite{Menche10,Dorogovtsev10,Litvak13,Raschke10}. To overcome
this problem, the Spearman's rank correlation coefficient
$\rho$ for degrees has been proposed for measuring assortative
(or disassortative) mixing \cite{Litvak13}. This quantity is
defined by
\begin{equation}
\rho=\frac{\displaystyle 2\sum_{(i,j)\in E}R_{k_{i}}R_{k_{j}}-\frac{1}{2M}\left[\sum_{(i,j)\in E}(R_{k_{i}}+R_{k_{j}})\right]^{2}}
  {\displaystyle \sum_{(i,j)\in E}(R_{k_{i}}^{2}+R_{k_{j}}^{2})-\frac{1}{2M}\left[\sum_{(i,j)\in E}(R_{k_{i}}+R_{k_{j}})\right]^{2}},
\label{eq:rho}
\end{equation}
where $R_{k_{i}}$ is the rank of the degree of node $i$ and the
meaning of the summation is the same as in
Eq.~(\ref{eq_assortativity}). From the above definition, it is
clear that $\rho$ is the Pearson's correlation coefficient for
degree ranks.

There are several ways to determine the rank $R_{k}$ if plural
nodes in a network have the same degree. To avoid this problem,
Litvak \textit{et~al}. \cite{Litvak13} resolve the rank
degeneracy for the same degree nodes by using random numbers.
This method, however, makes it difficult to calculate
analytically $\rho$ even for a deterministic network such as
the $(u,v)$-flower. An alternative way of ranking is to rank
degrees of $2M$ end nodes in ascending order with assigning the
average rank of degenerated degrees to them \cite{Zhang16}. In
this case, the rank of an end node with degree $k$ is given by
\begin{equation}
R_{k}=\frac{kN_{k}+1}{2}+\sum_{k'=0}^{k-1}k'N_{k'} ,
\label{eq:rank}
\end{equation}
where $N_{k}=NP(k)$ is the number of nodes with degree $k$.
Using this ranking, the Spearman's rank correlation coefficient
$\rho$ can be written as
\begin{equation}
\rho=\frac{\displaystyle 2\sum_{(i,j)\in E}R_{k_{i}}R_{k_{j}}-\frac{M}{2}(2M+1)^{2}}
  {\displaystyle \sum_{(i,j)\in E}(R_{k_{i}}^{2}+R_{k_{j}}^{2})-\frac{M}{2}(2M+1)^{2}},
\label{eq:rho2}
\end{equation}
where we use $\sum_{(i,j)\in E}(R_{k_{i}}+R_{k_{j}})=M(2M+1)$.
It should be noted that topologies of networks with a specific
degree sequence are reflected only in the first term
$\sum_{(i,j)\in E}R_{k_{i}}R_{k_{j}}$ of the numerator. Since
the rank $R_{k}$ one-to-one corresponds to $k$, analytical
calculations of $\rho$ are possible for some deterministic
networks. For example, $\rho$ for the $n$th generation
$(u,v)$-flower $G_{n}^{(u,v)}$ with $u\ge 2$ and $n\ge 2$ is
presented by (see the Appendix)
\begin{equation}
\rho_{n}^{(u,v)}=-\frac{z\left(z^{2}+z+1\right)}{1-z^{3(n-1)}} ,
\label{eq:uv-rho}
\end{equation}
where $z\equiv 2/(u+v)$ is less than or equal to $1/2$. For an
infinitely large $(u,v)$-flower ($n\to \infty$),
$\rho_{\infty}^{(u,v)}=-z(z^{2}+z+1)$ is negative for any $z$
as expected from Fig.~\ref{fig:1}(a), while the assortativity
$r$ for $G_{\infty}^{(u,v)}$ with $z\ge 1/4$ becomes zero as
shown in the Appendix.

We construct a MD network $G_{\text{MD}}$ by rewiring edges of
$G_{\text{ini}}$ so as to minimize the Spearman's rank
correlation coefficient $\rho$. The actual rewiring procedure
is as follows:

(i) Choose randomly two edges $(i_{1},j_{1})$ and
$(i_{2},j_{2})$ with four different end nodes.

(ii) Rewire these edges to $(i_{1},j_{2})$ and $(i_{2},j_{1})$,
if this re\-wiring does not make multiple edges and not increase
the Spearman's rank correlation coefficient $\rho$.

(iii) Repeat (i) and (ii) until $\rho$ cannot be decreased
anymore by rewiring.

This disassortative rewiring (DR) also does not change the
degree sequence. Therefore, $N_{k_{m}}$ of $G_{\text{MD}}$ is
still given by Eq.~(\ref{uv_Nkm}). For some degree sequences,
the rank correlation $\rho$ cannot reach its minimum value by
the above DR scheme because of local minimum trap. In such a
case, an optimization algorithm based on simulated annealing
\cite{Donetti05} must be employed instead of the present DR.
However, the above simple DR can minimize $\rho$ at least for
the degree sequence specified by Eq.~(\ref{uv_Nkm}), as
mentioned later.

\section{RESULTS}
\label{sec:3}
\begin{figure}[tttt]
\begin{center}
\includegraphics[width=0.45\textwidth]{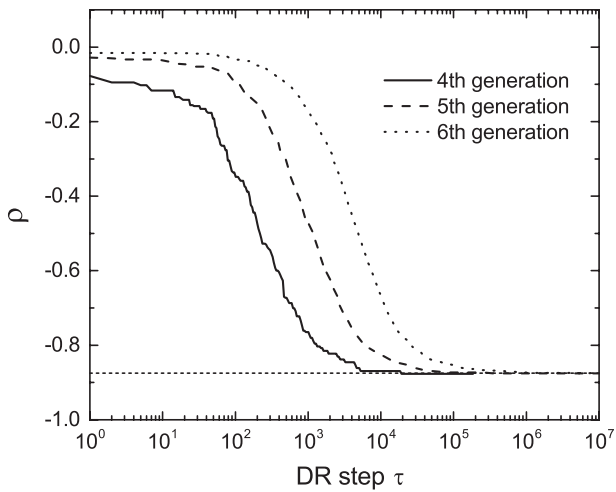}
\caption{Rank correlation coefficient $\rho$ as a function of
the number of DR steps $\tau$ starting from the initial
networks ($G_{\text{ini}}$) formed by $5M$ operations of RR for
the 4th (solid line), 5th (dashed line), and 6th (dotted line)
generation $(2,2)$-flowers ($G_{0}$), where $M$ is the number
of edges in the networks. The horizontal dotted line indicates
the value of $\rho_{6}^{(2,2)}$.}
\label{fig:2}
\end{center}
\vspace{-0.5cm}
\end{figure}
Figure \ref{fig:2} shows how the rank correlation coefficient
$\rho$ decreases with the DR step $\tau$ starting from
$G_{\text{ini}}$. Initial networks $G_{\text{ini}}$ are formed
by $5M$ operations of RR for $(2,2)$-flowers ($G_{0}$), where
$M$ is the number of edges. Solid, dashed, and dotted curves in
Fig.~\ref{fig:2} represent the results for the 4th, 5th, and
6th generation $(2,2)$-flowers as $G_{0}$. The numbers of nodes
$N$ and edges $M$ are $N=172$, $684$, and $2\,732$ and $M=256$,
$1\,024$,  and $4\,096$ for the 4th, 5th, and 6th generations,
respectively. The initial values of $\rho$ at $\tau=0$ (i.e.,
$\rho$ for $G_{\text{ini}}$) are slightly negative even though
$G_{\text{ini}}$ is randomly rewired enough times. This is
because self-loops and multiple edges are not allowed in
$G_{\text{ini}}$. Since the probability to have such edges by
RR with allowing them decreases with the network size, $\rho$
for $G_{\text{ini}}$ becomes close to zero as the generation
increases. The rank correlations $\rho$ monotonically decrease
with the DR step, and reach negative constant values.
These convergence values coincide with $\rho_{n}^{(2,2)}$
presented by Eq.~(\ref{eq:uv-rho}), namely,
$\rho_{4}^{(2,2)}=-0.8767$, $\rho_{5}^{(2,2)}=-0.8752$, and
$\rho_{6}^{(2,2)}=-0.8750$ (indicated by the horizontal dotted
line in Fig.~\ref{fig:2}). The rank correlations never diminish
from these values even if repeating DR operation many times.

We can show that the above convergence value $\rho_{n}^{(u,v)}$
gives the minimum $\rho$ among networks with the degree
sequence specified by Eq.~(\ref{uv_Nkm}), that is, networks
with $\rho_{n}^{(u,v)}$ are nothing but MD networks. For
simplicity, let us discuss the case of $\rho_{n}^{(2,2)}$ and
define two sets, $S_{N_{k}}$ and $S_{\text{MD}}$, of networks
derived from the $n$th generation $(2,2)$-flower. The set
$S_{N_{k}}$ includes all networks with the degree sequence
specified by Eq.~(\ref{uv_Nkm}), while $S_{\text{MD}}$ is a
proper subset of $S_{N_{k}}$ in which every network is composed
of edges connecting two nodes with the lowest degree $k_{1}=2$
and a higher degree $k_{m}=2^{m}$ with $m\ge 2$. Namely, a
network in $S_{\text{MD}}$ has the joint probability $P(k,k')$
for $k< k'$ given by
\begin{equation}
P(k,k')=\frac{1}{2M}\sum_{m=2}^{n}k_{m}N_{k_{m}}\delta_{k2}\delta_{k'k_{m}} ,
\label{eq:Pkk'}
\end{equation}
where $N_{k_{m}}$ is presented by Eq.~(\ref{uv_Nkm}) and
$M=w^{n}$ is the number of edges. The rank correlation
coefficient $\rho$ takes a constant value for networks in
$S_{\text{MD}}$ because $\rho$ of Eq.~(\ref{eq:rho2}) is
uniquely determined by $P(k,k')$ and $R_{k}$ and the rank
$R_{k}$ defined by Eq.~(\ref{eq:rank}) does not change for
networks in $S_{N_{k}}$. Since the $(2,2)$-flower is an element
of $S_{\text{MD}}$, this constant value is equal to
$\rho_{n}^{(2,2)}$. If a network with the degree sequence
$N_{k_{m}}$ given by Eq.~(\ref{uv_Nkm}) has a joint probability
$P(k,k')$ different from Eq.~(\ref{eq:Pkk'}), the network
possesses edges connecting lowest degree nodes to each other.
In order to prove that $\rho$ of such a network is larger than
$\rho_{n}^{(2,2)}$, let us consider a network $\tilde{G}$
formed by rewiring two edges $(i,j)$ and $(i',j')$ of a network
$G \in S_{\text{MD}}$ to $(i,i')$ and $(j,j')$. Here, the
degrees of these end nodes are $k_{i}=k_{i'}=2$ and $k_{j}\ge
k_{j'}>2$. The network $\tilde{G}$ has the same $R_{k}$ as $G$
because $G,\tilde{G}\in S_{N_{k}}$, while $P(k,k')$ for
$\tilde{G}$ is slightly different from Eq.~(\ref{eq:Pkk'}). For
both $G$ and $\tilde{G}$, we have $R_{k_{i}}=R_{k_{i'}}$ and
$R_{k_{j}}\ge R_{k_{j'}}>R_{k_{i}}$, because $R_{k}$ is a
monotonically increasing function of $k$. These relations for
$R_{k}$ give the inequality,
\begin{equation*}
R_{k_{i}}R_{k_{i'}}+R_{k_{j}}R_{k_{j'}} > R_{k_{i}}R_{k_{j}}+R_{k_{i'}}R_{k_{j'}} .
\end{equation*}
The left-hand side of the above inequality represents the
contribution to the first term in the numerator of
Eq.~(\ref{eq:rho2}) from the edges $(i,i')$ and $(j,j')$ of
$\tilde{G}$, while the right-hand side is that from the edges
$(i,j)$ and $(i',j')$ of $G$. Other edges of $G$ and
$\tilde{G}$ give the same contribution to this term. The
remaining terms of Eq.~(\ref{eq:rho2}) do not change between
$G$ and $\tilde{G}$. Therefore, the above inequality shows that
the rank correlation $\rho_{G}=\rho_{n}^{(2,2)}$ for the
network $G$ is less than $\rho_{\tilde{G}}$ for $\tilde{G}$.
This implies that $\rho_{n}^{(2,2)}$ provides the minimum value
$\rho_{\text{MD}}$ of $\rho$ for networks in $S_{N_{k}}$. In
other words, $S_{\text{MD}}$ is the set of MD networks
($G_{\text{MD}}$) within $S_{N_{k}}$. In the case of
Fig.~\ref{fig:2}, $\rho$ reaches $\rho_{\text{MD}}$ at
$\tau=18\,473$, $286\,415$, and $4\,762\,744$ for the 4th, 5th,
and 6th generations, respectively. Further DR operations after
these steps realize different topology networks
$G_{\text{MD}}$'s in $S_{\text{MD}}$. The above argument for
$\rho_{n}^{(2,2)}$ can be easily generalized to
$\rho_{n}^{(u,v)}$.

\begin{figure}[tttt]
\begin{center}
\includegraphics[width=0.48\textwidth]{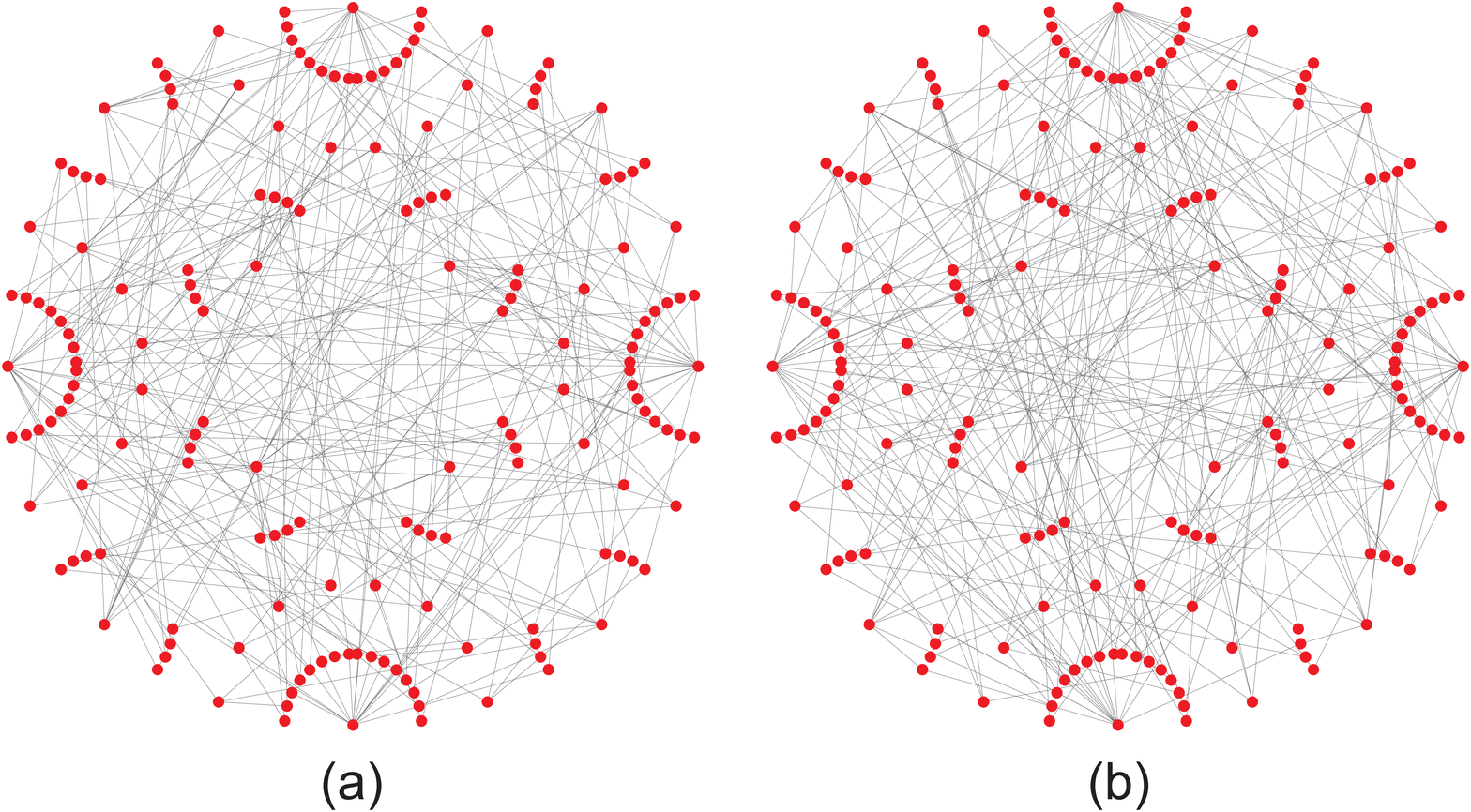}
\caption{Networks after (a) $3\times 10^{5}$ and
(b) $6\times 10^{5}$ operations of DR from the network shown
by Fig.~\ref{fig:1}(b).}
\label{fig:3}
\end{center}
\end{figure}
Examples of MD networks formed by DR's from $G_{\text{ini}}$
shown in Fig.~\ref{fig:1}(b) are presented in Fig.~\ref{fig:3}.
These networks seem to be very different from the $4$th
generation $(2,2)$-flower ($G_{0}$) shown in
Fig.~\ref{fig:1}(a), though the nearest-neighbor degree
correlations are the same as $G_{0}$. In order to quantify such
differences in network topology, we employ two indices. One is
the spectral distance $U_{GG'}$ between networks $G$ and $G'$
with the same size $N$ \cite{Wilson08,Gu15}. The distance
$U_{GG'}$ is defined by
\begin{equation}
U_{GG'}=\sqrt{\sum_{\alpha=1}^{N}\left[\left(\mu_{\alpha}^{G}-\mu_{\alpha}^{G'}\right)^{2}+
\left(\nu_{\alpha}^{G}-\nu_{\alpha}^{G'}\right)^{2}\right]} .
\label{eq:spec_dist}
\end{equation}
Here, $\mu_{\alpha}^{G}$ and $\nu_{\alpha}^{G}$ are the
$\alpha$th eigenvalues (in ascending order) of the adjacency
matrix $A=[a_{ij}]$ and the Laplacian matrix $L=[l_{ij}]$,
respectively, where $a_{ij}=1$ if the nodes $i$ and $j$ are
connected in $G$, $a_{ij}=0$ otherwise, and
$l_{ij}=-a_{ij}+\delta_{ij}\sum_{j'}a_{ij'}$. Since $U_{GG'}$
is invariant under the similarity transform, $U_{GG'}=0$ for
networks $G$ and $G'$ being isomorphic to each other. It should
be remarked, however, that two cospectral networks providing
$U_{GG'}=0$ are not always isomorphic. Although $U_{GG'}$
becomes large when the topology of $G'$ largely deviates from
$G$, it is not clear what type of topological differences
strongly affects $U_{GG'}$. Thus, we introduce another measure,
the correlation distance $V_{GG'}$, to quantify the topological
difference between networks. The quantity $V_{GG'}$ is defined
as
\begin{equation}
V_{GG'}=\frac{1}{2}\sum_{k,k',l}\left|P_{G}(k,k',l)-P_{G'}(k,k',l)\right| ,
\label{eq:corr_dist}
\end{equation}
where $P_{G}(k,k',l)$ is the joint probability that randomly
chosen two nodes have the degrees $k$ and $k'$ and are
separated by the shortest path distance $l$ to each other.
Similar to $U_{GG'}$, $V_{GG'}$ takes the minimum value $0$ if
$G=G'$. $V_{GG'}$ becomes maximum if two distribution functions
$P_{G}(k,k',l)$ and $P_{G'}(k,k',l)$ have no overlap, and is\break
bounded by $1$ because of the normalization condition\break
$\sum_{k,k',l} P_{G}(k,k',l)=1$. A large correlation distance
$V_{GG'}$ implies that $G'$ displays very different
(long-range) degree correlations from $G$.

\begin{figure}[tttt]
\begin{center}
\includegraphics[width=0.48\textwidth]{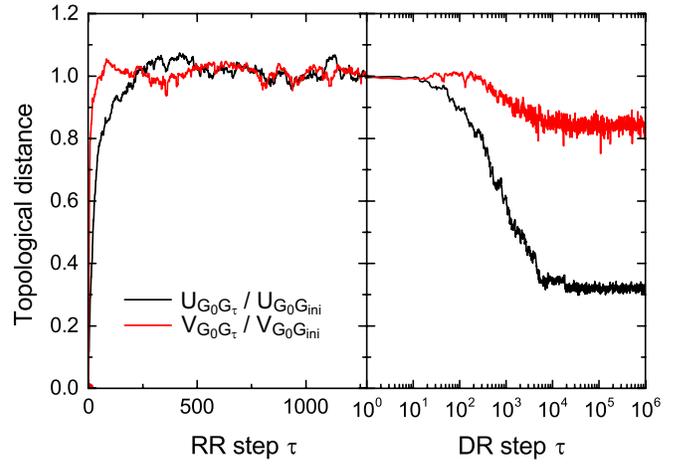}
\caption{Spectral distance $U_{G_{0}G_{\tau}}$
(black line) and correlation distance $V_{G_{0}G_{\tau}}$ (red line)
rescaled by $U_{G_{0}G_{\text{ini}}}$ and $V_{G_{0}G_{\text{ini}}}$,
respectively, where $G_{0}$ is the 4th generation $(2,2)$-flower,
and $G_{\text{ini}}$ is the network after $5M=1\,280$ operations of
RR from $G_{0}$. The left panel displays the distances
during $5M$ operations of RR starting from $G_{0}$, while
the right panel shows the distance changes by DR's starting from
$G_{\text{ini}}$.}
\label{fig:4}
\end{center}
\end{figure}
We examined how largely MD networks are different from the
$(u,v)$-flower by computing the above topological distances.
The left panel of Fig.~\ref{fig:4} shows $U_{G_{0}G_{\tau}}$
and $V_{G_{0}G_{\tau}}$ as a function of the number of RR steps
$\tau$, where $G_{0}$ is the original 4th generation
$(2,2)$-flower $G_{4}^{(2,2)}$ and $G_{\tau}$ is the network
after $\tau$ operations of RR from $G_{0}$. The right panel
depicts the same quantities, but $\tau$ is the number of DR
steps starting from $G_{\text{ini}}$, and $G_{\tau}$ is the
network after $\tau$ operations of DR from $G_{\text{ini}}$. In
this panel, data for $\tau\ge 18\,473$ represent the
topological distances of MD networks from $G_{4}^{(2,2)}$. The
topological distances must become zero at some values of $\tau$
in this region because the $G_{4}^{(2,2)}$ included in
$S_{\text{MD}}$ is reachable from any MD network by DR's.
Nevertheless, both the spectral and correlation distances never
drop to zero, at least within the present window of $\tau$.
This implies that the topology of MD network in $S_{\text{MD}}$
is diverse and most of MD networks have very different
structures from the $(u,v)$-flower.

\begin{figure}[tttt]
\begin{center}
\includegraphics[width=0.48\textwidth]{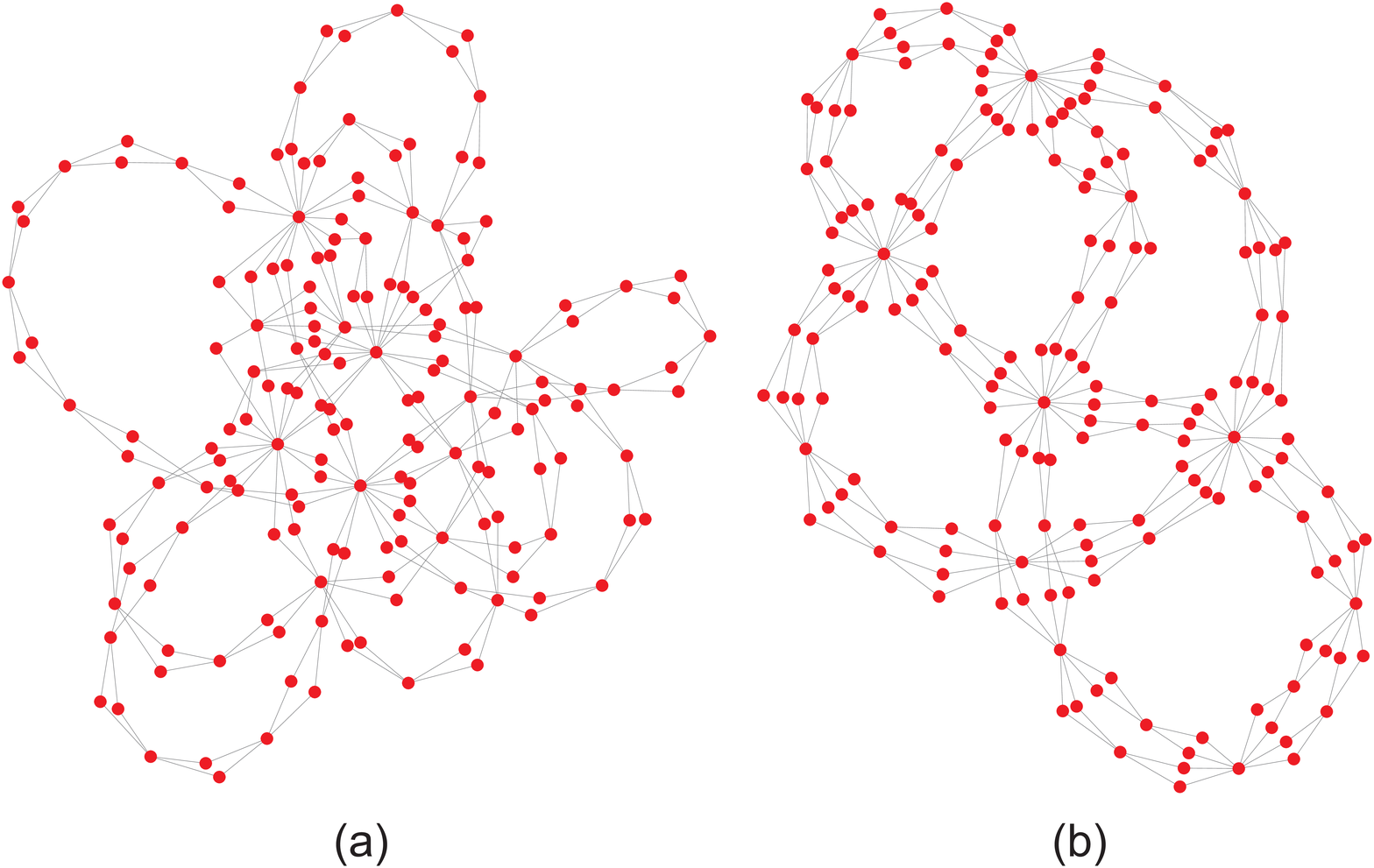}
\caption{Networks after (a) $320$ and (b) $80$ operations
of UR in units of $G_{1}^{(2,2)}$ and
$G_{2}^{(2,2)}$ starting from $G_{4}^{(2,2)}$, respectively.}
\label{fig:5}
\end{center}
\vspace{-0.5cm}
\end{figure}

The diverse topologies of MD networks can be readily understood
through the idea of \textit{unit rewiring}. The $n$th
generation $(u,v)$-flower $G_{n}^{(u,v)}$ is composed of
$w^{n-m}$ pieces of $G_{m}^{(u,v)}$ ($m\le n$). If we regard
$G_{m}^{(u,v)}$ as a superedge of $G_{n}^{(u,v)}$,
$G_{n}^{(u,v)}$ is equivalent to $G_{n-m}^{(u,v)}$ with
superedges. Let us perform RR operations for $G_{n-m}^{(u,v)}$
with superedges, which corresponds to RR's in units of the
subgraph $G_{m}^{(u,v)}$ in $G_{n}^{(u,v)}$. We call such a
rewiring a unit rewiring (UR). As examples, networks after
$320$ and $80$ operations of UR in units of $G_{1}^{(2,2)}$ and
$G_{2}^{(2,2)}$ starting from $G_{4}^{(2,2)}$ are depicted in
Figs.~\ref{fig:5}(a) and 5(b), respectively. It should be
emphasized that the network after UR's has the same $P(k,k')$,
and thus $\rho$, as the original network $G_{n}^{(u,v)}$.
Therefore, all networks formed by UR's are elements of
$S_{\text{MD}}$. The number of networks formed by UR operations
in units of $G_{m}^{(u,v)}$ starting from $G_{n}^{(u,v)}$ is
equivalent to the number of networks with the same degree
sequence as $G_{n-m}^{(u,v)}$. Since even this number is quite
large \cite{Barvinok13}, the number of elements in
$S_{\text{MD}}$ that includes unit-rewired networks as a part
of it rises astronomically.

\begin{figure}[tttt]
\begin{center}
\includegraphics[width=0.43\textwidth]{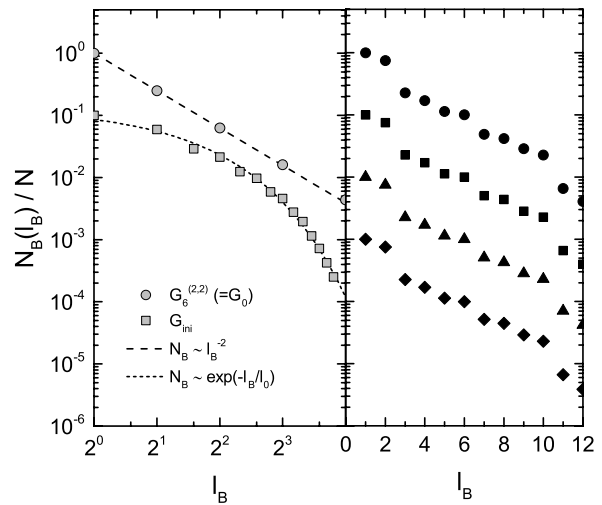}
\caption{Number of covering subgraphs $N_{\text{B}}$ as a
function of the subgraph diameter $l_{\text{B}}$ for the
original 6th generation $(u,v)$-flower $G_{6}^{(2,2)}$ (gray
circles in the left panel), $G_{\text{ini}}$ formed by RR's
from $G_{6}^{(2,2)}$ (gray squares in the left panel), and four
topologically different networks in $S_{\text{MD}}$ formed by
DR's from $G_{\text{ini}}$ (black symbols in the right panel).
The results are vertically shifted for graphical reasons.
Lines in the left panel are guides to the eye.}
\label{fig:6}
\end{center}
\end{figure}

Considering the property of unit rewiring, in a network
$G_{\text{UR}}$ formed by UR operations in units of
$G_{m}^{(u,v)}$, the degree-degree correlation or other
structural correlations in units of $G_{m}^{(u,v)}$ cannot
extend beyond the scale of them. This implies that the network
$G_{\text{UR}}$ ($\in S_{\text{MD}}$) does not possess the
fractal property as a long-range structural correlation. Thus,
we can conclude that disassortative degree mixing does not
always make a scale-free network fractal. Although the
$(u,v)$-flower in $S_{\text{MD}}$ has surely a fractal
structure, it can be demonstrated that fractal networks are
rather rare in the set $S_{\text{MD}}$. In the right panel of
Fig.~\ref{fig:6}, the number of covering subgraphs
$N_{\text{B}}$ [see Eqs.~(\ref{eq:def_sw}) and
(\ref{eq:def_fr})] is plotted as a function of $l_{\text{B}}$
for $4$ topologically very different networks in
$S_{\text{MD}}$, by employing the compact-box-burning algorithm
\cite{Song07}. These results show that
$N_{\text{B}}(l_{\text{B}})$ decreases exponentially with
$l_{\text{B}}$. In addition to these $4$ examples, we examined
$N_{\text{B}}(l_{\text{B}})$ for totally $100$ MD networks with
different topologies to each other. Our results show that
$N_{\text{B}}(l_{\text{B}})$ for every network obeys
Eq.~(\ref{eq:def_sw}), which implies that most of MD networks
with the scale-free property are not fractal, but have the
small-world property.
\begin{figure}[tttt]
\begin{center}
\includegraphics[width=0.46\textwidth]{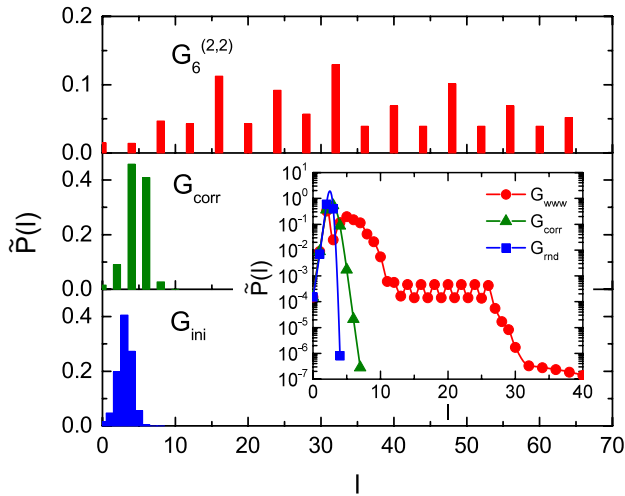}
\caption{Distance distribution $\tilde{P}(l)$
for the 6th generation $(2,2)$-flower $G_{6}^{(2,2)}$ (top),
networks ($G_{\text{corr}}$) formed by rewiring edges while
keeping $P(k,k')$ starting from $G_{6}^{(2,2)}$ (middle), and networks
($G_{\text{ini}}$) formed by RR's starting from $G_{6}^{(2,2)}$ (bottom).
The results for $G_{\text{corr}}$ and $G_{\text{ini}}$ are
averaged over $100$ connected samples. The inset shows
$\tilde{P}(l)$ for the WWW ($G_{\text{www}}$) with $N=325\,729$
nodes and $M=1\,497\,134$ edges \cite{URL_www} (red circles),
a network ($G_{\text{corr}}$) formed by rewiring edges while
keeping $P(k,k')$ starting from $G_{\text{www}}$ (green triangles), and
a network ($G_{\text{rnd}}$) formed by RR's starting from $G_{\text{www}}$
(blue squares). Curves are guides to the eye.}
\label{fig:7}
\end{center}
\end{figure}
Does the degree correlation have nothing to do with the fractal
property of a scale-free network? Our results demonstrate that
at least the nearest neighbor disassortative degree correlation
cannot be the origin of fractality in scale-free networks.
However, as seen in Fig.~\ref{fig:1}(a), the $(u,v)$-flower has
a distinguishing feature of a long-range repulsive correlation
between the same degrees. This is not found in a typical MD
network shown in Fig.~\ref{fig:3}. In particular, the
long-range repulsion between hub nodes is found not only in the
$(u,v)$-flower but also in other synthetic scale-free fractal
networks \cite{Song06}. In order to quantify the repulsive
correlation between hubs, we introduce the conditional
probability $P(l|k,k')$ of randomly chosen two nodes with
degrees $k$ and $k'$ being separated by the distance $l$ to
each other. If setting $k=k'=k_{\text{max}}$, this probability
indicates the distribution of shortest path distance among
nodes with the largest degree $k_{\text{max}}$ and gives
information on the hub-hub repulsion. To improve the
statistical reliability of this distribution, we define the
distance distribution function
$\tilde{P}(l)=c\sum_{k,k'}^{'}P(l|k,k')P(k)P(k')$, where $P(k)$
is the degree distribution, $\sum_{k,k'}^{'}$ represents the
summation over degrees of the top 2\% of high degree nodes, and
$c$ is the normalization constant. As expected, $\tilde{P}(l)$
for the $(u,v)$-flower shown in the top panel of
Fig.~\ref{fig:7} is distributed in a wide range of $l$, which
implies that high degree nodes are likely to be largely
separated from each other. The middle panel of Fig.~\ref{fig:7}
indicates $\tilde{P}(l)$ for networks ($G_{\text{corr}}$)
formed by rewiring many edges while keeping $P(k,k')$ starting
from the $(u,v)$-flower, where we should note that
$G_{\text{corr}}$ is a MD network. The actual re\-wiring process
can be done as follows: First we choose randomly two edges
$(i_{1},j_{1})$ and $(i_{2},j_{2})$ with four different end
nodes, where the degree of node $i_{1}$ is equal to the degree
of node $i_{2}$, then rewire them to $(i_{1},j_{2})$ and
$(i_{2},j_{1})$. The function $\tilde{P}(l)$ for
$G_{\text{corr}}$ is distributed in a narrow range of small
$l$. The width of the distribution is, however, slightly wider
than that for $G_{\text{ini}}$ (bottom panel of
Fig.~\ref{fig:7}) formed by RR's starting from the
$(u,v)$-flower. This is because higher degree nodes are never
directly connected in $G_{\text{corr}}$. In addition to the
$(u,v)$-flower, we examined $\tilde{P}(l)$ for the WWW
\cite{URL_www} as a real-world scale-free fractal network. This
network exhibits disassortative degree mixing ($\rho=-0.11$)
and is known to have the fractal dimension $D=4.41$
\cite{Kawasaki10,Song05}. As shown in the inset of
Fig.~\ref{fig:7}, $\tilde{P}(l)$ for the WWW has a long tail.
If the network is rewired while keeping $P(k,k')$,
$\tilde{P}(l)$ becomes much narrower than that for the original
WWW as seen by triangles in the inset and the network loses
fractality. These results suggest that the fractal property of
a scale-free network requires a long-range repulsive
correlation between similar degree nodes, particularly hub
nodes.

\section{CONCLUSIONS}
\label{sec:4}
We have studied the relation between fractality of scale-free
networks and their degree correlations. Real-world and
synthetic fractal scale-free networks are known to exhibit
disassortative degree mixing in common.  It is, however, not
obvious whether a negative correlation between nearest neighbor
degrees causes the fractal property of a scale-free network,
though the possibility of disassortativity being the origin of
fractality is suggested. In order to clarify this point, we
examined maximally disassortative (MD) networks prepared by
rewiring edges while keeping the degree sequence. For the
preparation of MD networks, uncorrelated networks are first
formed by rewiring randomly edges of the $(u,v)$-flower. Then,
these networks are again rewired to minimize the rank
correlation coefficient $\rho$. Our results show that there
exist a huge number of MD networks with different topologies
but most of them are not fractal. Therefore, it is concluded
that negative correlations between nearest neighbor degrees
cannot be the origin of fractality of scale-free networks. This
can be readily understood if we consider MD networks formed by
unit rewiring operations starting from the $(u,v)$-flower,
because unit-rewired networks cannot have any structural
correlations beyond the scale of the rewiring unit. In
addition, we studied the long-range repulsion between hub nodes
in fractal scale-free networks and their rewired networks while
keeping the joint probability $P(k,k')$. The results for the
$(u,v)$-flower and a real-world fractal scale-free network show
that distances between large degree nodes in fractal scale-free
networks are much longer than those in their rewired networks.
This fact prompts the speculation that fractality of scale-free
networks requires a long-range repulsive correlation in similar
degrees.

We should point out that networks treated in this work are a
bit special. As mentioned in Sec.~\ref{sec:1}, MD networks are
generally composed of communities in each of which all nodes
with a specific low degree and higher degree nodes are
connected alternately. This gives MD networks a long-range
degree correlation. On the other hand, an MD network with the
same degree sequence as the $(u,v)$-flower ($u>1$) consists of
only one community, because the total number of spokes from the
lowest degree nodes is equal to or larger than that from all
remaining higher degree nodes. We then cannot expect the
long-range degree correlation in MD networks with the same
degree sequence as the $(u,v)$-flower. Nevertheless, our
conclusion is considered to be still valid even for more
general networks. This is because disassortative degree mixing
does not introduce any long-range repulsive correlations
between similar degree nodes in a community of a general MD
network. Therefore, the MD network is not fractal at least
below the scale of the community.

Our speculation about the long-range repulsive correlation
between similar degree nodes for fractal scale-free networks
should be checked by further investigations. For this purpose,
it is significant to define a new index characterizing the
strength of such a repulsive correlation. It is interesting to
identify whether a network formed by rewiring edges to maximize
this index becomes fractal.

\begin{acknowledgements}
The authors thank S.~Tomozoe for fruitful discussions. This
work was supported by a Grant-in-Aid for Scientific Research
(No.~16K05466) from the Japan Society for the Promotion of
Science.
\end{acknowledgements}

\appendix*
\section{$\rho$ OF THE $(u,v)$-FLOWER}
Here, we derive Eq.~(\ref{eq:uv-rho}) for the $n$th generation
$(u,v)$-flower with $u\ge 2$ and $n\ge 2$. At first, we
determine the rank $R_{k_{m}}$ for the degree $k_{m}=2^{m}$
($m=1,2,\cdots,n$) in the $(u,v)$-flower. For the
$(u,v)$-flower, Eq.~(\ref{eq:rank}) can be written as
\begin{equation}
R_{k_{m}}=\frac{S(k_{m})+1}{2}+\sum_{m'=1}^{m-1}S(k_{m'}) ,
\label{eq:Rkm}
\end{equation}
where $S(k_{m})=k_{m}N_{k_{m}}$ is the total number of spokes
from nodes with degree $k_{m}$. Since $N_{k_{m}}$ is given by
Eq.~(\ref{uv_Nkm}), we have
\begin{equation}
S(k_{m})=
\begin{cases}
2^{m}(w-2)w^{n-m} & (1\le m< n) , \\
2^{n}w & (m=n) ,
\end{cases}
\label{eq:Skm}
\end{equation}
where $w=u+v$. Therefore, the rank of degree $k_{m}$ is
presented by
\begin{equation}
R_{k_{m}}=
\begin{cases}
\displaystyle
w^{n}\left[2-\biggl(\frac{w}{2}+1\biggr)\biggl(\frac{2}{w}\biggr)^{m}\right] + \frac{1}{2} & (1\le m< n) , \\[10pt]
\displaystyle
2w^{n}-2^{n-1}w+\frac{1}{2} & (m=n) .
\end{cases}
\label{eq:Rkm2}
\end{equation}

To calculate $\rho$ by Eq.~(\ref{eq:rho2}), we need to evaluate
\begin{equation}
X=\sum_{(i,j)\in E}R_{k_{i}}R_{k_{j}} ,
\label{eq:X}
\end{equation}
and
\begin{equation}
Y=\sum_{(i,j)\in E}\left(R_{k_{i}}^{2}+R_{k_{j}}^{2}\right) ,
\label{eq:Y}
\end{equation}
for the $(u,v)$-flower. Considering that nodes with degree
$k_{m}$ for $m\ge 2$ always connect to the lowest degree nodes
with degree $k_{1}=2$, the number of edges whose end nodes have
the degree ranks $R_{k_{m}}\ (m\ge 2)$ and $R_{k_{1}}$ is
$S(k_{m})$. The remaining
$\left[S(k_{1})-\sum_{m=2}^{n}S(k_{m})\right]/2$ edges connect
the lowest degree nodes to each other. Thus, the quantity $X$
is presented by
\begin{equation}
X=\sum_{m=2}^{n}S(k_{m})R_{k_{m}}R_{k_{1}}+
\frac{1}{2}\left[S(k_{1})-   \sum_{m=2}^{n}S(k_{m}) \right]R_{k_{1}}^{2}.
\label{eq:X_2}
\end{equation}
Using Eqs.~(\ref{eq:Skm}) and (\ref{eq:Rkm2}), the quantity $X$
is calculated as
\begin{equation}
X=\left(1-z^{2}\right)M^{3}+M^{2}+\frac{1}{4}M ,
\label{eq:X_3}
\end{equation}
where $z=2/w$ and the number of edges $M$ is given by $(2/z)^{n}$.
The summation over edges in Eq.~(\ref{eq:Y}) is also rewritten as
\begin{equation}
Y = \sum_{m=1}^{n} S(k_{m})R_{k_{m}}^{2},
\label{eq:Y_2}
\end{equation}
and $Y$ can be calculated as
\begin{equation}
Y=\frac{2(z+1)^{2}}{z^{2}+z+1}M^{3}+2M^{2}+\frac{1}{2}M-\frac{2z}{z^{2}+z+1}\frac{2^{3n}}{z^{3}}.
\label{eq:Y_3}
\end{equation}
The calculations of
Eqs.~(\ref{eq:X_3}) and (\ref{eq:Y_3}) become easier if we
utilize the obvious relations $\sum_{m=1}^{n}S(k_{m})=2M$ and
$\sum_{m=1}^{n}S(k_{m})R_{k_{m}}=M(2M+1)$. From
Eqs.~(\ref{eq:rho2}), (\ref{eq:X_3}) and (\ref{eq:Y_3}), the
rank correlation coefficient $\rho_{n}^{(u,v)}$ for the $n$th
generation $(u,v)$-flower is then calculated as
\begin{equation}
\rho_{n}^{(u,v)}=-\frac{z\left(z^{2}+z+1\right)}{1-z^{3(n-1)}} ,
\label{eq:uv-rho_app}
\end{equation}
which is identical to Eq.~(\ref{eq:uv-rho}). For the
$(2,2)$-flower, for example, $\rho_{n}^{(2,2)}$ is calculated
as
\begin{equation}
\rho_{n}^{(2,2)}=-\frac{7}{8}\left[1-\frac{1}{2^{3(n-1)}}\right]^{-1} ,
\label{rho_for_22-flower}
\end{equation}
which gives $\rho_{\infty}=-7/8$ for $n\to \infty$.

The assortativity $r$ defined by Eq.~(\ref{eq_assortativity})
for the $n$th generation $(u,v)$-flower with $u\ge 2$ and $n\ge
2$ can be calculated by a similar way. The result is given by
\begin{equation}
r=
\begin{cases}
\displaystyle
-\frac{2^{n+2}z^{n}-a^{2}(2z-1)-4}{(2z-1)(a^{2}-b)} & (w\ne 4,8) ,\\[15pt]
\displaystyle
-\frac{(n+1)^{2}-4n}{(3\cdot 2^{n}-2)-(n+1)^{2}} & (w=4) ,\\[15pt]
\displaystyle
-\frac{4^{n-1}-2^{n}+1}{4^{n-1}(3n-8)+3\cdot 2^{n}-1} & (w=8) ,
\end{cases}
\label{eq:r_uv}
\end{equation}
where
\begin{equation}
a= \frac{(2z)^{n}+2z-2}{2z-1} ,
\label{eq:a}
\end{equation}
and
\begin{equation}
b= \frac{3\cdot 4^{n}z^{n}+4z-4}{4z-1} .
\label{eq:b}
\end{equation}
For $n\to \infty$, the assortativity $r_{\infty}$ converges as
\begin{equation}
r_{\infty}=
\begin{cases}
0 & (w\le 8) ,\\
\displaystyle
-\frac{2}{w(w-2)} & (w\ge 9) .
\end{cases}
\label{eq:r_inf_uv}
\end{equation}
Since the scale-free exponent $\gamma$ given by
Eq.~(\ref{uv_gamma}) is less than $4$ for $w\le 8$, the above
result is consistent with the general fact
\cite{Menche10,Dorogovtsev10,Litvak13} that $r$ cannot be
negative for infinitely large scale-free networks with
$\gamma\le 4$.

\end{document}